\documentclass [12pt,a4paper]{article}

\usepackage[utf8]{inputenc}
\usepackage[english]{babel}
\usepackage[top=2cm,bottom=2cm,left=2.5cm,right=2.5cm,footskip=25pt]{geometry}
\usepackage{amsmath}
\usepackage{amssymb}
\usepackage{amsthm}
\usepackage{changepage}
\usepackage{comment}
\usepackage{xcolor}
\usepackage[bookmarks=false,hypertexnames=true]{hyperref} 
\usepackage[numbers,square,comma]{natbib} 
\newcommand{\quotes}[1]{``#1''}
\newcommand{\psibar}{{{\psi}^*}}
\newcommand{\totderiv}[2]{\frac{\dint #1}{\dint #2}}

\newcommand{\dint}{\mathop{}\!\mathrm{d}}

\newcommand{\flowavg}[1]{\overline{#1}}
\newcommand{\flowavgh}[2]{\overline{#1}^{#2}}
\newcommand{\diff}[1]{\delta{#1}}

\newtheorem{definition}{Definition}
\newtheorem{theorem}{Theorem}
\newtheorem{lemma}{Lemma}
\newtheorem*{remark}{Remark}
\begin{document}

\begin{adjustwidth}{1.8cm}{1.8cm}
    \centerline{\bf \Large From Klein-Gordon-Wave to Schr\"odinger-Wave:}
    \vspace{4mm}
    \centerline{\bf \Large a Normal Form Approach}
    \vspace{5mm}
    \centerline{\today} 
    \vspace{10mm}
    
    \noindent {\bf \large Gaia Marangon$^{1,*}$, Antonio Ponno$^{1,2}$, Lorenzo Zanelli$^{1,2}$}
    
    \vspace{2mm}
    \noindent $^{1}$ Department of Mathematics ``Tullio Levi-Civita", University of Padova, Via Trieste, 63, Padova, 35131, Italy.\\
    \noindent $^{2}$ Padua Quantum Technologies Research Center (QTech), Via Gradenigo, 6/A, Padova, 35131, Italy.\\
    \noindent $^*$ Corresponding author.
    
    \vspace{2mm}
    \noindent E-mails: \textit{marangon@math.unipd.it, ponno@math.unipd.it,\\ lzanelli@math.unipd.it}
    
    \vspace{5mm}
    \noindent {\bf Abstract. }
We consider a Klein-Gordon-Wave system, describing the evolution of a massive field and a massless one interacting through a Yukawa-like coupling, and we explicitly derive its Hamiltonian normal form to first and second order.
To the first-order approximation, the normal form results in a Schr\"odinger-Wave system,
 which reduces to the Schr\"odinger-Poisson one in the singular limit of vanishing perturbative parameter. The second-order approximation provides the successive corrections to the Schr\"odinger-Wave system, and is presented in order to show that higher-order approximations to all orders can be obtained by iterating our constructive procedure. 
The normal form technique adopted here formally extends the standard Birkhoff normal form procedure for harmonic oscillators to include a set of free particles in the unperturbed problem.
The mathematical result obtained here might explain, for example, the \quotes{cooling} process of ultra-light dark matter, the approximate validity of the Schr\"odinger-Poisson system describing its dynamics and the long term conservation of the total dark matter mass.    
    
    \vspace{5mm}
    Schr\"odinger-Wave, Schr\"odinger-Poisson, Dark Matter
    
\end{adjustwidth}


\section{Introduction} 
\label{sec:introduction}

In the past decades, wide efforts have been devoted to the derivation and the analysis of classical semi-relativistic or non-relativistic approximations of the Einstein-Klein-Gordon equations. Such approximate systems model the evolution of a scalar matter field interacting with a scalar gravitational field via Lagrangian terms of degree three, quadratic in the massive field and linear in the massless, gravitational one. We refer to such a kind of interaction as to a Yukawa-like coupling, since it is akin to the standard Yukawa coupling modeling the interaction of a spinor field with a real scalar field (in our case the spinor field is replaced by a real, scalar one) \citep{Buchbinder2021,Griffiths2008,Glimm1968}.
The differences in the above mentioned approximate models lie in the specific assumptions made for the two fields, such as the inclusion of a massive term for the gravitational field or the adoption of an overall Lorentz invariant structure.
Moreover, different assumptions may result in extremely different dynamical properties, including local or global well-posedness, stability of equilibria and number of conservation laws. 

However, most of these systems display a Hamiltonian structure, which allows for the application of specific tools to explore the possible connections among them. In particular, if the system at hand is in the form of a perturbation of an integrable case, one can apply the methods of Hamiltonian perturbation theory in order to build up the so-called normal form Hamiltonian \citep{AKN2006}, i.e. a Hamiltonian that, up to a certain change of variable, is equivalent to the original one but displays a simpler form to any given order. 
In such a way, neglecting small remainder terms one obtains an approximate system displaying the most relevant properties of the dynamics on a certain, long time-scale.

In the present work, we study the connections among three well known Hamiltonian models, 
namely the Klein-Gordon-Wave system, the Schr\"odinger-Wave system, and the Schr\"odinger-Poisson one. We show that, in a certain regime, depending on the value of a single dimensionless parameter and on the choice of the initial conditions, the Klein-Gordon-Wave dynamics is well approximated, in the normal form sense, by the that of a Schr\"odinger-Wave system. In turn, a large class of solutions of the latter system are close to those of the Schr\"odinger-Poisson one. 

We now describe in a heuristic way how the three systems connect, giving an intuitive anticipation of the results detailed in the following Sections.
Then, we comment on their possible physical interpretation at an informal level, referring to Appendix A for a more accurate description. A discussion on the literature follows. Formal details are deferred to specific Sections.

\medskip
 
Our starting point is a Klein-Gordon-Wave (KGW) model assuming a Klein-Gordon dynamics for the massive field $u(t,x)$ and a wave dynamics for the massless field 
$\phi(t,x)$ with Yukawa-like coupling, namely
\begin{align}
\begin{cases}
\Box u = \mu^2 u + 2\phi u \\
\Box \phi = u^2 
\end{cases} 
\ \ , \ \ 
\Box := \Delta-\partial_t^2\,.
\label{eq:KG-W}
\end{align} 
As shown later, the system admits a Hamiltonian variational formulation, the Yukawa-coupling term being given by $\int u^2\phi\dint{^3}x$.
System \eqref{eq:KG-W} is written in normalized, dimensionless form for the real valued fields $u$ and $\phi$ defined on $\mathbb{R}^4$, and its dynamics is ruled by the dimensionless parameter $\mu>0$, containing all the physical information on the system. Besides the boundary conditions ($u$ and $\phi$ finite at the origin and vanishing at infinity), the system is completed by the initial normalization condition
$\|u(0)\|^2:=\int_{\mathbb{R}^3} u^2(0,x)\dint{^3x}=1$, 
which is interpreted, as usual, as fixing the total number of particles at time $t=0$ (number which is arbitrarily rescaled to one). Such a condition is not preserved by the dynamics of the system, as expected in a Lorentz invariant model. Now, depending on the value of $\mu$, the dynamics of the KGW system \eqref{eq:KG-W} displays quite different features. For example, if one looks for stationary, i.e. time independent solutions, the system at hand reduces to a stationary Schr\"odinger-Poisson one, namely 
\begin{align}
\begin{cases}
 -\Delta u + 2\phi u = -\mu^2 u \\
\Delta \phi = u^2 
\end{cases}\ .
\label{eq:KGWstat}
\end{align} 
The latter system, with the condition $\|u\|^2=1$, is known to display a countably infinite number of spherically symmetric solutions corresponding to a positive, monotonically decreasing sequence of eigenvalues 
$\mu_j$ converging to $0$ \cite{Lions1980}. 
If the parameter $\mu$ is larger than $\mu_0$, such stationary states no longer exist,  and for intermediate values of $\mu$ nothing \quotes{simple} can be said. One is thus naturally led to study the dynamics of the KGW for large values of the parameter $\mu$, say $\mu\gg\mu_0$, where the KGW dynamics somehow simplifies, and quasi-stationary states appear again. Indeed, dividing the equations \eqref{eq:KG-W} by $\mu^2$, and defining the rescaled time $\tau=\mu t$, we get
\begin{align}
\begin{cases}
 \partial_\tau^2u+u=\frac{1}{\mu^2}(\Delta u-2\phi u) \\
\partial_\tau^2\phi=\frac{1}{\mu^2}(\Delta\phi-u^2)
\end{cases}\ .
\label{eq:KGWres}
\end{align} 
Now, it is clear that for large values of $\mu$, and initial conditions 
$(u,\phi,\partial_\tau u,\partial_\tau\phi)|_{t=0}$ of order one in $\mu$, the dynamics of
\eqref{eq:KGWres} stays close, on a certain time-scale to be determined, to the unperturbed one, consisting of a harmonic motion of $u$, and a linear growth of $\phi$ with time. Such a motion can be visualized as the product of a harmonic, or circular motion and a rectilinear one. Thus, at variance with the usual problems of perturbation theory, the unperturbed dynamics does not take place on a torus, but on a cylinder. In order to treat the right hand side of 
\eqref{eq:KGWres} as a perturbation, the small parameter of the theory being $1/\mu^2$, when $\mu^2\gg1$, it is convenient to introduce the complex variables 
\begin{equation}
\label{eq:Psi}
\Psi=\frac{u+i\partial_\tau u}{\sqrt{2}}\ e^{i\tau}\ \ ;\ \ 
\Psi^*=\frac{u-i\partial_\tau u}{\sqrt{2}}\ e^{-i\tau}
\end{equation}
where $i$ denotes the imaginary unit. Getting $u$ from \eqref{eq:Psi}, 
substituting it into \eqref{eq:KGWres}, and then defining a new rescaled time 
$T=\tau/\mu^2=t/\mu$, one gets
the following system
\begin{align}
\begin{cases}
i\partial_T\Psi = -\frac{1}{2}\Delta\Psi  + \phi \Psi +
\left[e^{i 2\mu^2T}\left(-\frac{1}{2}\Delta+\phi\right)\Psi^*\right] \\
\frac{1}{\mu^2}\partial^2_T\phi= \Delta\phi-|\Psi|^2 +\left[-\frac{1}{2}
\Psi^2e^{-i2\mu^2T}-\frac{1}{2}(\Psi^*)^2e^{i2\mu^2T}\right]
\end{cases}\ .
\label{eq:SWrem}
\end{align}  
We now observe that the two exponential factors appearing in the remainder terms in square brackets have a vanishing time-average over the extremely short period 
$\pi/\mu^2$. By time-averaging the right hand side of \eqref{eq:SWrem} over such fast oscillations, we are left with the Schr\"odinger-Wave (SW) system
\begin{align}
\begin{cases}
i\partial_T\Psi = -\frac{1}{2}\Delta\Psi  + \phi \Psi \\
\frac{1}{\mu^2}\partial^2_T\phi= \Delta\phi-|\Psi|^2
\end{cases}\ .
\label{eq:SW}
\end{align}    
This system is no longer Lorentz invariant. However, due to its gauge invariance, i.e. invariance under $\Psi\to e^{i\theta}\Psi$, the N\"other charge
$\|\Psi(T)\|^2=\int|\Psi(x,T)|^2\dint{^3}x$ is now a constant of motion. This is not surprising, since such a norm, when expressed in terms of $u$ and its time derivative, is nothing but the energy of the unperturbed harmonic motion. On the other hand, expressing $u$ in terms of $\Psi$ and $\Psi^*$, gives the constancy of $\|u\|^2$ on time average over the fast oscillations. We interpret such a result as the (approximate) conservation of the number of particles in the non relativistic regime corresponding to large values of the parameter 
$\mu$. The actual value of $\|\Psi\|^2$ can be set of order one as that of $\|u(0)\|^2$. 

In the singular limit $\mu^2\to\infty$, the SW system \eqref{eq:SW} reduces to the Schr\"odinger-Poisson (SP) model
\begin{align}
\begin{cases}
i\partial_T\Psi = -\frac{1}{2}\Delta\Psi  + \phi \Psi \\
\Delta \phi = |\Psi|^2 
\end{cases}\ .
\label{eq:SP}
\end{align} 
Of course, neglecting the term $\partial_T^2\phi/\mu^2$ in the second of equations \eqref{eq:SW} is a non trivial approximation, worth to be discussed. To such a purpose, 
we only observe here that systems \eqref{eq:SW} and \eqref{eq:SP} share the same
stationary (actually time-periodic) solutions of the form
$\Psi(T,x)=e^{-\imath\omega T}\chi(x)$ (determined, as usual, by fixing the norm $\|\chi\|^2=1$, for example). For solutions close to such stationary states the delay, or radiation effect, in $\phi$ is expected to be small due to the pre-factor $1/\mu^2$ in front of the time derivative. Observe that $\mu$ plays the same formal role of the speed of light in the wave equation for $\phi$. Thus, the limit $\mu\to\infty$ is actually a no radiation limit. A more refined argument is provided below, in Sec. \ref{sec:firstHamEq}.

\medskip

From a physical point of view, the systems \eqref{eq:KG-W}, \eqref{eq:SW} and \eqref{eq:SP} may certainly be relevant in different fields. We point out here a particularly interesting application to the problem of ultralight scalar dark matter dynamics at the galactic level; see e.g. \cite{Hui2021} and references therein. In such a context, the parameter $\mu$ appearing in \eqref{eq:KG-W} and encoding all the physics of the problem has the explicit expression
\begin{equation}
\label{eq:mu}
\mu=\frac{1}{N}\left(\frac{m_P}{m}\right)^2\ ,
\end{equation}
where $m$ is the mass of the candidate ultralight scalar dark matter particle (a spinless, chargeless boson), $N$ is the total mass of such particles, and $m_P=\sqrt{\hbar c/G}$ is the Planck mass ($\hbar$, $c$ and $G$ being the Planck constant, the speed of light and the gravitational constant, respectively). The value of $N$ is fixed by imposing that
$Nm$ be of the order of the total mass of dark matter in a given galaxy, say a few units of the baryon component. A short \quotes{derivation} of the KGW system \eqref{eq:KG-W} for this problem leading to the expression \eqref{eq:mu} for the parameter is reported in the Appendix \ref{sec:app_phys} at the end of the paper. We stress that the parameter \eqref{eq:mu} and its role in ruling the behaviour of self gravitating bosons, described by a SP system, has been introduced first by \citet{Ruffini1969}. The value of $m$ for scalar bosons has been first conjectured by \citet{Baldeschi1983} to be about $10^{-58}$ to $10^{-57}$ grams. With such a small value, and $N$ such that $Nm$ is of the order of 
$10^{12}$ solar masses, for example, we get a value of the parameter $\mu^2$ of the order $10^4$ to $10^6$, motivating our assumption of large $\mu$.
A possible physical implication of our result, namely approximating the KGW dynamics by the SW or the SP one, is a justification of the \quotes{cooling} mechanism for dark matter. Indeed, according to the standard cosmological model, dark matter has to be \quotes{cold}, i.e. non relativistic. Given the extraordinarily small value of their mass, necessary to form structures at the galactic level, the non relativistic behaviour of dark particles seems to be an interesting conclusion. The main mechanism presented above can be rephrased in terms of loss of Lorentz invariance, placed at the onset as a fundamental requirement of any field theory, and gain of gauge invariance, which is necessary to preserve the mass of the dark halo (the integral of $|\Psi|^2$). The latter conservation law is preserved to any finite order in our perturbative procedure (this is better explained below), implying that dark halos, if any, may persist on very long time-scales. 

\medskip


We now shortly report on the mathematical physics literature concerning the systems treated in the present paper.
 
The KGW system \eqref{eq:KG-W} emerges as a particular case of a larger class of problems; see the remark below. These problems were introduced by \citet{Ionescu2019} as a simplified version of the Einstein–Klein–Gordon equations of general relativity, with the aim of studying their global nonlinear stability. The latter was first proved in the case of small, smooth and compactly supported perturbations using the hyperboloidal foliation method \citep{LeFloch2016,LeFloch2023} (see also \citep{LeFloch2014} for a detailed description of the method) and then extended with different techniques to relax the compact support hypothesis \citep{Ionescu2019,Ionescu2020r}. Later works focused on specific configurations of this class of problems \citep{Dong2020,Ouyang2023,Lei2023}. This series of works provided a solid foundation for analyzing the same results for the more general Einstein-Klein-Gordon system \citep{Ionescu2020r,LeFloch2017,Ionescu2020,Wang2020}. 

The Schr\"odinger-Wave (SW) system \eqref{eq:SW}, mostly studied in the literature for $\varepsilon=1$, is a particular case of the more general Schr\"odinger-Klein-Gordon (SKG) model, which assumes the gravitational field $\phi(t,x)$ to be massive.
The analytical properties of the Schr\"odinger-Klein-Gordon system have been thoroughly analyzed. The global well posedness of the associated Cauchy problem has been established under different regularity conditions in a series of works \citep{Fukuda1978,Baillon1978,Hayashi1987,Bachelot1984,Colliander2008,Pecher2012,Akahori2005},
including the existence and characterization of wave solutions in the Schr\"odinger-Wave case of a massless gravitational field \citep{Ginibre2002}.
\citet{Ohta1996} first focused on the orbital stability of the ground standing wave solution, a result that was then thoroughly discussed in a series of works for the massive gravitational field case \citep{Kikuchi2008,Kikuchi2010,Kikuchi2011}. 

The wide interest in the Schr\"odinger-Klein-Gordon and Schr\"odinger-Wave models is motivated by their connection with the Nelson model, a consistent quantum field theory constructed by \citet{Nelson1964} that describes the Yukawa-like interaction of a non-relativistic nucleon field with a relativistic meson field. Ever since the introduction of the Nelson model, the SKG system is considered the classical limit of this theory, a result later proved by \citet{Ammari2016}. Based on preliminary studies \citep{Falconi2013,Ammari2014}, the latter authors manipulate the classical Schr\"odinger-Klein-Gordon system through Hamiltonian normal form techniques: using the flow of a suitably defined generating Hamiltonians as change of variables, they rewrite the system in a form that is suitable for quantization, avoiding - even in infinite-dimensional spaces - the incurrence of divergences and thus the necessity of cutoffs.

The SP system \eqref{eq:SP} was first suggested by \citet{Ruffini1969} as the weak-field, non-relativistic limit of the general relativistic Einstein-Klein-Gordon equations, approximating the behavior of a self-gravitating system of bosons. Subsequently, it has been used in a variety of contexts \citep{Paredes2020}, such as the description of boson stars \citep{Schunck2003} and ultralight scalar dark matter configurations \citep{Matos2024}, the analysis of the fundamentals of quantum mechanics \citep{Diosi1984,Penrose1998} and quantum gravity \citep{Penrose2014,Bahrami2014}, or the modeling of dilute cold atomic BECs \citep{Lahaye2009} and nonlinear optics setups \citep{Grossardt2016,Bekenstein2015}.
Its mathematical properties have been thoroughly analyzed, including the existence and uniqueness of spherically symmetric stationary states \citep{Lieb1977,Lions1980}, their numerical characterization \citep{Bernstein1998,Moroz1998,Harrison2003,Marangon2025JMP,Marangon2025PLA}, the smoothness and boundedness of the eigenfunctions \citep{Tod1999}, bounds for the eigenvalues \citep{Tod2001} and long-range asymptotic properties \citep{Kiessling2021}.\\

\begin{remark}
    The KGW system \eqref{eq:KG-W} analyzed in this work can be considered as a particular case of the following class of problems, which was introduced by \citet{Ionescu2019} based on the work by \citet{LeFloch2016}:
\begin{align*}
\begin{cases}
    (-\Box +1)u =\phi B^{\alpha\beta} \partial_\alpha\partial_\beta u + E\phi u\\
    -\Box \phi = A^{\alpha\beta}\partial_\alpha u \,\partial_\beta u +D u^2
\end{cases}\,,
\end{align*}
with $u$ representing the massive scalar field, $\phi$ replacing the deviation of the Lorentzian metric from the Minkowski metric and $A^{\alpha\beta},B^{\alpha\beta},D,E$ real constants. This system is obtained by expressing the Einstein–Klein–Gordon equations of general relativity in harmonic gauge, and by keeping only, schematically, quadratic interactions that involve the massive scalar field \citep{Ionescu2019}. In this work, we specialize to $A^{\alpha\beta}=B^{\alpha\beta}=0$ and adjust the remaining constants to obtain the KGW system \eqref{eq:KG-W}.
\end{remark}
 

The rest of the paper is devoted to the exposition of the mathematical technique used to show that the SW system \eqref{eq:SW} approximates the KGW system \eqref{eq:KG-W} when $\mu^2\gg1$. At variance with the heuristic method exposed above, we rely on the theory of Hamiltonian normal forms \citep{Arnold1978,Gaeta2002,Cicogna1994}. In our case, we construct the normal form of the KGW Hamiltonian up to the second order in $\varepsilon=\mu^{-2}$, showing that its first-order truncation results in the SW Hamiltonian.
The second order truncation appears a bit cumbersome, and is provided in order to show that the SW system is just a first order approximation of the KGW one, and that we are able to explicitly compute the successive approximations, in principle, to any finite perturbative order. 

The standard normal form construction is based on the averaging principle (we make reference to its implementation in \citep{Gallone2022}), which works under the hypothesis that the unperturbed flow is bounded. Such a condition is satisfied, in particular, in Birkhoff-Gustavson problems \citep{Birkhoff1927,Gustavson1966}, consisting of perturbations of $N$ harmonic oscillators, whose Hamiltonian is $\sum_{k=1}^N \omega_kI_k$, with given frequencies $\omega_1,\dots,\omega_N$. The unperturbed dynamics preserves the actions $I_k$ and evolves the conjugated angles $\theta_k$ on a torus.

A possible extension of the above problem consists, for example, in adding $M$ free particles to the system, decoupled from the oscillators in the unperturbed problem, and moving on an unbounded domain. The resulting Hamiltonian reads $\sum_{k=1}^N \omega_kI_k + \sum_{j=1}^M \frac{p_j^2}{2}$, and the associated Hamiltonian dynamics preserves the actions $I_k$ and the momenta $p_j$ of the particles, evolving the angles $\theta_k$ on a torus while
the particle coordinates $q_j$ undergo a linear unbounded growth. This is a major problem in applying the standard techniques of \citep{Gallone2022}.  

Now, the unperturbed Hamiltonian of the KGW system \eqref{eq:KG-W} is just an infinite dimensional version of the Hamiltonian of oscillators and particles described above (see Sec. \ref{sec:harmOscFreeParti} for a detailed description). In the present paper we extend the Hamiltonian normal form construction to problems of this kind, building up a suitable perturbative scheme. In particular, we manage to consistently compute time averages only along the bounded component of the unperturbed flow, while keeping the free particle component fixed. Although our system is completely resonant ($\omega_k=1$ for all $k$), the same procedure can also be applied, with the due caution in treating small divisors, to more general cases.  

An important issue in Hamiltonian perturbation theory consists in the rigorous control of the remainders: their smallness should be estimated and, in neglecting them, the difference between the solutions of the original system and those of the truncated one should be also estimated. Such controls are non trivial even when dealing with finite-dimensional systems \citep{Vey1978,Ito1989,Zung2002,Zung2005,Stolovitch2000,Russmann1964,Bruno1971}. The extension to the infinite dimensional case is even more delicate, as it requires controlling additional divergences \citep{Kuksin2010,Grebert2014,Kappeler2003,Bambusi1998,Bambusi2002,Bambusi2005,Bambusi2006,Bambusi2020,Baldi2021}.
In the present paper, we provide only the formal part of the normal form construction for the Klein-Gordon-Wave problem, deferring the analysis of its well-posedness to a more technical work. Further comments on this point are reported in Sec. \ref{sec:harmOscFreeParti} below. 

\medskip

The paper is structured as follows. In Section \ref{sec:model} we present the KGW model \eqref{eq:KG-W}, describing its Hamiltonian structure and setting some preliminary definitions and properties.
In Section \ref{sec:mainResult} the main result is reported: the Hamiltonian normal form of the KGW system to first and second order is provided, and its truncation is discussed (Sections \ref{sec:firstHamEq}, \ref{sec:secondHamEq}), analyzing the resulting approximated systems with a specific focus on the limit recovering the SP system (Section \ref{sec:SWtoSP}). A discussion of the interpretation of the KGW system as a set of harmonic oscillators and free particles with higher-order coupling is also included (Section \ref{sec:harmOscFreeParti}). Section \ref{sec:normalFormComputation} is devoted to the explicit computation of the normal form, each step being first carried out in an algebraic form and then specialized to the KGW problem. Remarks on the extension of the procedure to higher orders are also included (Section \ref{sec:higherOrders}). 
Section \ref{sec:discussionConclusions} contains the conclusions. Finally, there are two Appendices. In Appendix A we provide a \quotes{derivation} of the KGW system describing the dynamics of scalar dark matter. In Appendix B we report the explicit computation of a quantity related to the second order normal form.

\section{KGW model and perturbative setting}
\label{sec:model}

We start our analysis by rewriting equations \eqref{eq:KG-W}, in the rescaled form \eqref{eq:KGWres}, here reported for convenience: 
\begin{align}
\begin{cases}
\partial_\tau^2 u + u = \varepsilon(\Delta u - 2\phi u) \\
\partial_\tau^2 \phi = \varepsilon(\Delta \phi - u^2) 
\end{cases}\ ;\ \ \varepsilon:=\frac{1}{\mu^2}\ .
\label{eq:timeScaling1}
\end{align}
We recall that $\tau=\mu t$, $t$ being the original time variable. 
The KGW equations above display a perturbative structure: in the limit $\varepsilon \to 0$ - henceforth referred to as the unperturbed limit - the Klein-Gordon equation reduces to the harmonic oscillator equation $\partial_\tau^2 u + u=0$, while the wave equation becomes the free-particle equation $\partial_\tau^2 \phi =0$. The Laplacian terms and the interaction terms appear as corrections of order $\varepsilon$.

As can be easily checked, equations \eqref{eq:timeScaling1} are the Hamilton equations - written in second order form - associated to the Hamiltonian
\begin{align}
\label{eq:Heps}
H_{\varepsilon} &= \underbrace{\int \frac{p_u^2 + u^2 + p_{\phi}^2}{2}}_{H_0} +\ \varepsilon \underbrace{\int \left(\frac{|\nabla u|^2 + |\nabla \phi|^2}{2} + \phi u^2\right)}_{H_1}\ . 
\end{align}
Here we highlighted the unpertubed Hamiltonian $H_0$ and its perturbation $H_1$; the field variables $p_u$ and $p_\phi$ denote the momenta conjugated to
$u$ and $\phi$, respectively.  
Observe that the unperturbed component $H_0$ includes two decoupled contributions: a harmonic oscillator Hamiltonian $h$ in the massive field coordinates and a free particle Hamiltonian $k$ in the massless field coordinates:
\begin{align}
    H_0 &= h + k\,, &&
    h:=\int \frac{p_u^2 + u^2}{2}\,, &&
    k:=\int \frac{p_{\phi}^2}{2}\,, &
    &\{h,k\}=0\ ,
    \label{eq:H0structure}
\end{align}
where $\{\cdot ,\cdot\}$ denotes the Poisson bracket. 
Concerning the perturbation $H_1$, it displays a polynomial structure, involving the term 
$\int|\nabla u|^2/2$, quadratic in the massive field, the term $\int|\nabla\phi|^2/2$, quadratic in the massless field, and the Yukawa coupling $\int u^2\phi$, quadratic in the massive field and linear in the massless one. 

We now perform a further change of variables on the massive field coordinates, which is suited for harmonic oscillators. We observe that, while convenient for computations, this change of variables is not strictly necessary, since the whole normal form construction is actually coordinate-free. The only assumptions, explained in detail in Section \ref{sec:mainResult}, concern the underlying structure of the unperturbed Hamiltonian $H_0$, described in \eqref{eq:H0structure}, which is independent of the specific choice of the coordinates. 
We thus introduce the complex field $\psi$ given by
\begin{equation}
\psi := \frac{u + ip_u}{\sqrt{2}}\ \ ;\ \ \psibar = \frac{u - ip_u}{\sqrt{2}}\ ,
\end{equation}
namely the infinite-dimensional analogue of the complex Birkhoff coordinates for the harmonic oscillator \citep{Birkhoff1927}.
The Hamiltonial \eqref{eq:Heps}, in the new variables, reads
\begin{align}
H_{\varepsilon} &= \underbrace{\int \left[|\psi|^2 + \frac{p_{\phi}^2}{2}\right]}_{H_0} +\  \varepsilon \underbrace{\int \left[\frac{|\nabla\psi+\nabla\psibar|^2}{4} + \frac{|\nabla\phi|^2}{2} + \frac{\phi(\psi+\psibar)^2}{2}\right]}_{H_1} \,,
\label{eq:HpertuIniziale}
\end{align}
where now $h= \int |\psi|^2$ and  $k=\int \frac{p_{\phi}^2}{2}$. Observe that the canonical Poisson bracket $\{u(x),p_u(y)\}=\delta(x-y)$ implies now that 
$\{\psi(x),\psibar(y)\} = -i \delta(x-y)$.

In terms of the coordinates $\psi$ and $\psi^*$ the unperturbed equations read
\begin{align}
\begin{cases}
    \partial_t \psi &= -i\psi\\
    \partial_t \psibar &= i\psibar\\
    \partial_t\phi &= p_\phi\\
    \partial_t p_\phi &= 0
\end{cases}\ .
\end{align}
On the other hand, the flows associated to $h$ and $k$ are easily obtained, namely
\begin{align}
\begin{aligned}
    \Phi_h^s (\psi,\psibar,\phi,p_{\phi})&= (e^{-is}\psi,\,e^{is}\psibar,\phi,p_{\phi})\ ; \\
    \Phi_k^s(\psi,\psibar,\phi,p_{\phi}) &= (\psi,\psibar,\phi+p_{\phi}s,\,p_{\phi})\ ,
\end{aligned}
\end{align}
showing that the flow of the unperturbed Hamiltonian $H_0=h+k$ is actually the composition of the flows along the two commuting components $h$ and $k$: $\Phi_{H_0}^s =\Phi_{h+k}^s= \Phi_h^s\circ \Phi_k^s$. 

Observe that the unperturbed flow $\Phi_h^s$ of the harmonic oscillator Hamiltonian $h$ is bounded, since it involves only a phase shift. In particular, the unperturbed flow 
preserves $h=\int|\psi|^2$. In contrast, the flow $\Phi_k^s$ of the Hamiltonian $k$ is unbounded, growing linearly with time

The presence of an unbounded component in the flow of the unperturbed Hamiltonian represents the main difference with respect to the ordinary Birkhoff problem, and the main obstacle to the application of the standard averaging principle technique.

\subsection{Preliminary Definitions and Properties}
Before stating our main result, let us report some standard definitions and properties that are common in Hamiltonian normal form techniques and that will be used in the following.

\begin{definition}[Lie derivative operator]
\label{def:lieDeriv}
    Let $K$ be an Hamiltonian, with $X_K$ the associate Hamiltonian vector field. The Lie derivative $L_K$ of a functional $F$ along the vector field $X_K$ is defined by  \citep{Arnold1978}:
    \begin{align*}
        L_K F := \{F,K\} = X_K \cdot \nabla F\,.
    \end{align*}
\end{definition}
Observe that with this definition, the operator $L_K$ inherits all the properties of the Poisson brackets, such as linearity ($L_{K_1+K_2}=L_{K_1}+L_{K_2}$) and antisymmetry ($L_{K_1}K_2 = -L_{K_2}K_1$). In addition, if two Hamiltonians commute, $\{K_1,K_2\}=0$, then the corresponding Lie derivative operators commute: $L_{K_1}L_{K_2}=L_{K_2}L_{K_1}$.

\begin{lemma}[Flow composition] 
\label{lem:flowComposition}
    The composition of a functional $F$ with the flow $\Phi_K^s$ of the Hamiltonian $K$ can be expressed as:
    \begin{align}
        F \circ \Phi_K^s = e^{sL_K}F = (1+L_K +\frac{1}{2}L_K^2+\dots)\,,
    \label{eq:flowComposition}
    \end{align}
    where the exponential form of the flow composition allows for a formal Taylor expansion.
\end{lemma}

\begin{definition}[Time average along the flow]
\label{def:timeAverages}
    The time average of a functional $F$ along the flow $\Phi_K^s$ of the Hamiltonian $K$ is defined as 
    \begin{align*}
        \flowavgh{F}{K}:= \lim_{t \to \infty}\frac{1}{t} \int_0^t F\circ\Phi_K^s \dint{s}=\lim_{t \to \infty}\frac{1}{t} \int_0^t e^{sL_K}F \dint{s}  \,.
    \end{align*} 
    For periodic flows $\Phi_K^s$ with fundamental period $T$, the definition is equivalent to averaging on the period: 
    \begin{align*}
        \flowavgh{F}{K}= \frac{1}{T} \int_0^T e^{sL_K}F \dint{s}\,.
    \end{align*}
\end{definition}
Observe that in this paper we will need to compute only time averages along the $2\pi$-periodic flow of the unperturbed Hamiltonian component $h$ defined in \eqref{eq:H0structure}. 
To simplify the notation, we will omit the superscript $h$, denoting only by $\flowavg{F}$ the time average along the flow $\Phi_h^s$:
\begin{align*}
        \flowavg{F} = \flowavgh{F}{h}= \frac{1}{2\pi} \int_0^{2\pi} e^{sL_h}F \dint{s}\,.
    \end{align*}
\begin{definition}[Deviation from the average]  Let $h$ be the Hamiltonian defined \eqref{eq:H0structure}. The difference of a functional $F$ with respect to its time average along the flow $\Phi_h^s$ is denoted by:
\begin{align}
    \diff{F} := F- \flowavg{F}
\end{align}
\end{definition}
\begin{lemma}
\label{lem:eqLhG}
    Let $h$ be the Hamiltonian defined in \eqref{eq:H0structure} and $L_h=\{\cdot,h\}$ its Lie derivative.  For any functional $F$, the solution of the equation:
    \begin{align*}
        L_h G = \diff{F} 
    \end{align*}
    is given by:
    \begin{align}
        G = \mathcal{G} + L_h^{-1} \diff{F} = \mathcal{G} + \frac{1}{2\pi}\int_0^{2\pi} s e^{s L_h}\diff{F}\,,
    \label{eq:defLh-1}
    \end{align}
    where $\mathcal{G}$ is an arbitrary element of $\ker{L_h}$, i.e. any function satisfying $\{\mathcal{G},h\}=0$.
\end{lemma}
The proof of this Lemma can be found in \citep{Gallone2022}.

\section{Main Result}
\label{sec:mainResult}
In this section we state the main result of this paper, namely the construction of the first- and second-order Hamiltonian normal forms for the KGW problem, and we study the Hamilton equations obtained by neglecting the remainders. 

As mentioned in Section \ref{sec:model}, the novelty of our problem lies in the presence of an unbounded component $\Phi_k^s$ in the unperturbed flow $\Phi_{H_0}^s$. In order to handle that, we first adapt the definition of normal form, as follows. 
\begin{definition}[Normal Form]
\label{def:normalForm}
Let $H_0=h+k$ be an Hamiltonian with two decoupled components, $\{h,k\}=0$, both integrable, with bounded flow for $h$ and unbounded flow for $k$. 
A Hamiltonian $H_\varepsilon$ is said to be in normal form to order $n$ with respect to $h$ if it is in the form
\begin{align*}
    H_\varepsilon = H_0 + \sum_{j=1}^n \varepsilon^j Z_j + \mathcal{R}_{n+1}
\end{align*}
where $\{Z_j,h\}=0$ for any $j = 1,\dots,n$, and $\mathcal{R}_{n+1}$ is of order $\varepsilon^{n+1}$.
\end{definition}
At variance with the usual definition of normal form, we here require that the perturbations
$Z_j$ Poisson-commute only with $h$, and not with the whole unperturbed Hamiltonian $H_0=h+k$. Observe that $\{H_0,h\}=0$ as well, due to the decoupling $\{h,k\}=0$, so that the commuting property characterizing the normal form holds even at order zero. 

\begin{remark}
An important consequence of the definition \ref{def:normalForm} of normal form is the following: the condition $\{Z_j,h\}=0$ implies that the truncated normal form Hamiltonian obtained by neglecting the remainder $\mathcal{R}_{n+1}$ will commute with $h$, i.e. $h=\int |\psi|^2$ is an integral of motion for the approximated dynamics to any finite order.
\end{remark}

We can now state the main result.
 \begin{theorem}[Normal Form for KGW]
 \label{thm:normalForm}
 There exist two generating Hamiltonians $G_1,G_2$ and a  canonical transformation 
 \begin{align*}
     \mathcal{C}^{-1} &:=\Phi_{G_1}^{\varepsilon}\circ \Phi_{G_{2}}^{\varepsilon^2} 
 \end{align*}
 mapping the Hamiltonian $H_\varepsilon$ of the KGW problem, defined in \eqref{eq:HpertuIniziale}, into a new Hamiltonian: 
 \begin{align}
     \widetilde{H}_{\varepsilon} = 
     H_\varepsilon \circ \mathcal{C}^{-1}
     = H_0 + \varepsilon Z_1 + \varepsilon^2 Z_2 + \mathcal{R}_3 
 \label{eq:hamNormForm_mainRes}
 \end{align}
 which is in normal form to order $2$ with respect to $h$, according to Definition \ref{def:normalForm}. Its components are defined by:
 \begin{align}
     Z_1 := \flowavg{H_1}\,, \quad
     Z_2 := \flowavg{F_2}\,, \quad
     F_2:= L_{G_1}H_1 +\frac{1}{2}L_{G_1}^2H_0\,,
 \label{eq:Zdef_mainRes}
 \end{align}
 with explicit expressions:
 \begin{align*}
     Z_1 =&  \int \left[ \frac{|\nabla\phi|^2}{2} + \frac{|\nabla\psi|^2}{2} + \phi|\psi|^2 \right]\,,\\
     Z_2 =& -\frac{1}{8} \int |\Delta \psi|^2 + 
    \frac{1}{4} \int \phi \left( \psibar \Delta \psi + \psi \Delta \psibar \right) +\\
    &+\frac{i}{16}\int p_\phi \left[\psibar \Delta\psi-\psi\Delta\psibar\right]
    - \frac{1}{2} \int \phi^2 |\psi|^2 + \frac{1}{16} \int |\psi|^4 \,.
 \end{align*}
 The generating Hamiltonians are defined by:
 \begin{align}
     G_1 &:= (1 + L_h^{-1}L_k)^{-1}L_h^{-1}\,\diff{H_1}\,,  
     \qquad
     \diff{H_1} :=H_1-\flowavg{H_1} \,,
     \label{eq:G1def_mainRes} \\
     G_2  &:= (1 + L_h^{-1}L_k)^{-1}L_h^{-1}\,\diff{F_2}  \,,
     \qquad
     \diff{F_2} :=F_2-\flowavg{F_2}\,,
    \label{eq:G2def_mainRes}
 \end{align}   
 with explicit expression for $G_1$:
 \begin{align*}
      G_1 =& \int \left[\frac{i}{8}((\nabla\psi)^2 - (\nabla\psibar)^2 ) + \frac{i}{4} \phi (\psi^2 - \psibar^2) + \frac{1}{8}p_\phi (\psi^2 + \psibar^2) \right] 
    \,.
 \end{align*}
 \end{theorem}
An explicit expression for $G_2$ is reported in Section \ref{sec:normalFormComputation}.  
Notice that $G_2$ is not involved in the explicit computation of $Z_2$, its expression becoming necessary only at higher orders. 
The proof of the theorem is given in Section \ref{sec:normalFormComputation}.

\subsection{First Order Hamilton Equations}
\label{sec:firstHamEq} \label{sec:SWtoSP}
We now truncate the normal form Hamiltonian \eqref{eq:hamNormForm_mainRes} at first-order, neglecting all contributions of order $\varepsilon^2$. We get:
\begin{align*}
    \widetilde{H}_{(1)} := H_0 + \varepsilon Z_1 
    = \left[ \int |\psi|^2 + \int \frac{p_\phi^2}{2} \right] + \varepsilon \left[ \int \frac{|\nabla \phi|^2}{2} + \frac{|\nabla \psi|^2}{2} + \phi |\psi|^2 \right] \,.
\end{align*}
The associated Hamilton equations read:
\begin{align*}
\begin{cases}
    i \partial_\tau\psi = \,\,\,\,\,\frac{\delta \widetilde{H}_{(1)}}{\delta \psibar} = \psi + \varepsilon \left[ -\frac{1}{2} \Delta \psi + \phi \psi \right] \\
    i \partial_\tau \psibar = -\frac{\delta \widetilde{H}_{(1)}}{\delta \psi} = \psibar + \varepsilon \left[- \frac{1}{2} \Delta \psibar + \phi \psibar \right] \\
    \partial_\tau \phi \,\,= \,\,\,\,\,\frac{\delta \widetilde{H}_{(1)}}{\delta p_\phi} = p_\phi \\
    \partial_\tau p_\phi= -\frac{\delta \widetilde{H}_{(1)}}{\delta \phi} = \varepsilon \left[ \Delta \phi - |\psi|^2 \right] 
\end{cases}\,.
\end{align*}
The second equation is (obviously) just the complex conjugate of the first one, so that we omit it. The $\psi$ term in the first equation is removed by time-dependent gauge transformation $\psi = e^{-i\tau} \Psi$, and the equations for $\phi$ and $p_\phi$ can be combined in second order form, to give
\begin{align*}
\begin{cases}
i \partial_\tau \Psi = \varepsilon \left[ -\frac{1}{2}\Delta \Psi + \phi \Psi \right] \\
\partial_\tau^2 \phi \,\,= \varepsilon \left[ \Delta \phi - |\Psi|^2 \right] 
\end{cases}\,.
\end{align*}
We now perform another time re-scaling, setting $T = \varepsilon \tau$, giving
\begin{align}
\begin{cases}
i \partial_T \Psi = -\frac{1}{2}\Delta \Psi + \phi \Psi  \\
\varepsilon\partial_T^2 \phi \,\,= \Delta \phi - |\Psi|^2 
\end{cases}\ ,
\label{eq:KGW_firstOrdHamEquas}
\end{align}
i.e. the SW system \eqref{eq:SW}, which was there obtained by a heuristic method of averaging over fast oscillations. As mentioned in the Introduction, in the singular limit $\varepsilon\to0$ the  SW system \eqref{eq:SW} reduces to the SP system \eqref{eq:SP}. We there gave a short argument 
justifying the possibility to consider the dynamics of certain solutions of the SP system close to that of the corresponding solutions of the SW system \eqref{eq:KGW_firstOrdHamEquas}. That intuition can be refined a bit, as follows. 
First, we define $\varphi:=\phi-\Delta^{-1}|\Psi|^2$, where $\Delta^{-1}|\Psi|^2$ denotes the solution of the Poisson equation $\Delta\phi=|\Psi|^2$. Then, by substituting 
$\phi=\varphi+\Delta^{-1}|\Psi|^2$ into \eqref{eq:KGW_firstOrdHamEquas}, we get
\begin{align}
\begin{cases}
i \partial_T \Psi = -\frac{1}{2}\Delta \Psi + (\Delta^{-1}|\Psi|^2) \Psi  +\varphi\Psi \\
\Box_\varepsilon\varphi= \varepsilon\Delta^{-1}\partial_T^2|\Psi|^2 
\end{cases}\ ;\ \Box_\varepsilon:=\Delta-\varepsilon\partial_T^2\ .
\label{eq:SWvarphi}
\end{align}
Now, formally solving for $\varphi$ by imposing causality (only retarded solution) and no homogeneous component, denoting such a solution by 
$\varphi=\varepsilon\Box_\varepsilon^{-1}\Delta^{-1}\partial_T^2|\Psi|^2$, and substituting it into the first equation, yields
\begin{equation}
i \partial_T \Psi = -\frac{1}{2}\Delta \Psi + (\Delta^{-1}|\Psi|^2) \Psi +
\varepsilon\left[\Box_\varepsilon^{-1}\Delta^{-1}\partial_T^2|\Psi|^2\right]\Psi\ .
\end{equation}
In this way, at least at a formal level, it turns out that the radiative correction to the SP equation is of order $\varepsilon$. Once again, we observe that on the stationary states, of the form $\Psi(T,x)=e^{-i\omega T}\chi(x)$, the radiative correction exactly vanishes, so that we expect that for initial conditions close enough to them, the radiative correction is actually small.

\subsection{Second Order Hamilton Equations}
\label{sec:secondHamEq}
We now truncate the normal form Hamiltonian \eqref{eq:hamNormForm_mainRes} at second-order, thus getting:
\begin{align*}
    \widetilde{H}_{(2)} =& H_0 + \varepsilon Z_1 + \varepsilon^2 Z_2 \\
    =& \left[ \int |\psi|^2 + \int \frac{p_\phi^2}{2} \right] + \varepsilon \left[ \int \frac{|\nabla \phi|^2}{2} + \frac{|\nabla \psi|^2}{2} + \phi |\psi|^2 \right] +\\
    & +\varepsilon^2 \left[ -\frac{1}{8} \int |\Delta \psi|^2 + 
    \frac{1}{4} \int \phi \left( \psibar \Delta \psi + \psi \Delta \psibar \right) +\right.\\
    & \left.
    +\frac{i}{16}\int p_\phi \left[\psibar \Delta\psi-\psi\Delta\psibar\right]
    - \frac{1}{2} \int \phi^2 |\psi|^2 + \frac{1}{16} \int |\psi|^4 \right] \,.
\end{align*}
The associated Hamilton equations read:
\begin{align*}
\begin{cases}
    i \partial_\tau \psi = \,\,\,\,\,\frac{\delta \widetilde{H}_{(2)}}{\delta \psibar} = \psi + \varepsilon \left[ \frac{1}{2} (-\Delta \psi) + \phi \psi \right] +\\
    \qquad\qquad\qquad\,\,\,\,\,\,\,\,\, +\varepsilon^2 \left[ -\frac{1}{8} \Delta (\Delta \psi) + 
    \frac{1}{4}\left( \phi (\Delta \psi) +  \Delta (\phi \psi) \right) + \right.\\
    \qquad\qquad\qquad\,\,\,\,\,\,\,\,\, \left.
    +\frac{i}{16}\left( p_\phi (\Delta \psi) -\Delta (p_\phi \psi) \right)
    - \frac{1}{2} \phi^2 \psi + \frac{2}{16} \psi^2 \psibar \right] \\
    \partial_\tau \phi \,\,= \,\,\,\,\,\frac{\delta \widetilde{H}_{(2)}}{\delta p_\phi} = p_\phi 
    +\varepsilon^2 \left[ 
    \frac{i}{16}(\psibar \Delta\psi-\psi\Delta\psibar)
    \right]\\
    \partial_\tau p_\phi= -\frac{\delta \widetilde{H}_{(2)}}{\delta \phi} = \varepsilon \left[ \Delta \phi - |\psi|^2 \right] + \varepsilon^2 \left[ 
    -\frac{1}{4}(\psibar \Delta \psi + \psi \Delta \psibar)
    +\phi |\psi|^2 \right]
\end{cases}\,.
\end{align*}
As in the first-order analysis, the $\psi$ term in the first equation is removed by the gauge transformation $\psi e^{-i\tau} \Psi$ and the time is rescaled to $T=\varepsilon \tau$, to give
\begin{align*}
\begin{cases}
    i \partial_T\Psi =  \left[ -\frac{\Delta \Psi}{2} + \phi \Psi \right] + \varepsilon \left[ -\frac{\Delta (\Delta \Psi)}{8} + 
    \frac{\phi (\Delta \Psi)+\Delta (\phi \Psi)}{4} 
    +\frac{i\left(p_\phi (\Delta \Psi)-\Delta (p_\phi \Psi) \right)}{16} 
    - \frac{\phi^2 \Psi}{2} + \frac{|\Psi|^2 \Psi}{8} \right] \\
    \varepsilon \partial_T \phi \,\,= p_\phi 
    +\varepsilon^2 \left[ 
    \frac{i(\Psi^* \Delta\Psi-\psi\Delta\Psi^*)}{16}
    \right]\\
    \partial_T p_\phi =  \left[ \Delta \phi - |\Psi|^2 \right] + \varepsilon \left[ 
    -\frac{\Psi^* \Delta \Psi + \Psi \Delta \Psi^*}{4}
    +\phi |\Psi|^2 \right]
\end{cases}\,.
\end{align*}
As before, the equations for $\phi$ and $p_\phi$ can be combined in second order form. which is performed by computing the time derivative of the equation for $\phi$. On the right hand side we get a $\partial_Tp_\phi$ contribution, and another contribution that is of order $\varepsilon^2$ and can be consistently neglected. Thus, to order $\varepsilon$, we get
\begin{align*}
\begin{cases}
i \partial_T \Psi =  \left[ -\frac{\Delta \Psi}{2} + \phi \Psi \right] + \varepsilon \left[ -\frac{\Delta (\Delta \Psi)}{8} + 
\frac{\phi (\Delta \Psi)+\Delta (\phi \Psi)}{4} 
+\frac{i\left(p_\phi (\Delta \Psi)-\Delta (p_\phi \Psi) \right)}{16} 
- \frac{\phi^2 \Psi}{2} + \frac{|\Psi|^2 \Psi}{8} \right] \\
\varepsilon \partial_T^2 \phi \,\,= \left[ \Delta \phi - |\Psi|^2 \right] + \,\varepsilon \left[ 
    -\frac{\Psi^* \Delta \Psi + \Psi \Delta \Psi^*}{4}
    +\phi |\Psi|^2 \right] 
\end{cases}\ .
\end{align*}

\subsection{Harmonic oscillators and free particles}
\label{sec:harmOscFreeParti}
In describing the KGW model in Section \ref{sec:model} we remarked that the unperturbed Hamiltonian $H_0$ is composed of a harmonic oscillator Hamiltonian $h$ and a free particle Hamiltonian $k$. To further clarify this statement, let us expand the fields $\psi,\psibar$ and $\phi,p_\phi$ on a given orthonormal basis 
$\{\varphi_k\}_{k=1}^\infty$ of $L^2(\mathbb{R}^3)$, namely
\begin{align}
    \psi(t,x) &= \sum_{k=1}^\infty z_k(t)\varphi_k(x)\ ;&
    \phi(t,x) &= \sum_{k=1}^\infty q_k(t) \varphi_k(x)\ ;&
    p_\phi(t,x) &= \sum_{k=1}^\infty p_k (t) \varphi_k(x)\ ,
\label{eq:KGW_basisFuncNotation}
\end{align} 
In these variables, the unperturbed Hamiltonian - still denoted $H_0$ with a slight abuse of notation - reads:
\begin{align*}
    H_0 = \sum_{k=1}^\infty |z_k|^2 + \sum_{k=1}^\infty \frac{p_k^2}{2}\,.
\end{align*}
With this notation, we can read the KGW Hamiltonian 
$H_\varepsilon=H_0+\varepsilon H_1(z,z^*,q)$ as defining an infinite set of harmonic oscillators, described by the Birkhoff variables $\{z_k,z_k^*\}_{k=1}^\infty$, and of free particles, described by the variables $\{q_k,p_k\}_{k=1}^\infty$, coupled to higher orders in the perturbative parameter $\varepsilon$ (the Birkhoff variables are connected to the action-angle ones by the relation $z_k=\sqrt{I_k}e^{-i\theta_k}$). In our case, all harmonic oscillators share the same frequency $\omega_k=1$ and all the free particles share the same mass $m_k=1$.

In this work we derive the normal form for the KGW in a formal way. 
A more detailed study would include a check on the analytical details of the computation, such as the existence of the flow of the generating Hamiltonians or the explicit dependence on $\varepsilon$ of the remainder $\mathcal{R}_3$. To such a purpose, the most convenient way out consists in writing the Hamiltonian $H_\varepsilon$ as above, truncating all the infinite sums, both in $H_0$ and in $H_1$ to a suitable cutoff, i.e. neglecting the contributions with index $k$ higher than a given threshold $N$. In such a way one is left with a finite-dimensional problem, presenting no serious obstacle in this case, since the unperturbed motion of the oscillators is completely resonant. The difficult part of this approach consists in recovering the original problem by letting $N \to \infty$. This procedure, introduced by \citet{Bambusi2005}, is extremely delicate. Its application to the present problem is in progress.

The method introduced in the present paper applies to more general perturbation problems of the form $K=K_0+K_1(z,z^*,q,p)$, where
\begin{align*}
    K_0 = \sum_{k=1}^\infty \omega_k|z_k|^2 + \sum_{k=1}^\infty \frac{p_k^2}{2m_k}\,.
\end{align*}
We stress that, in such an extension, small divisors may appear, depending on the specific conditions on the frequencies $\{\omega_k\}_{k=1}^\infty$ (see e.g. \citep{Yoccoz1992}). 
 

\section{Normal Form Computation}
\label{sec:normalFormComputation}
As prescribed by the standard averaging principle, computing the normal form to second-order requires looking for a canonical transformation in the form:
\begin{align}
    \mathcal{C}^{-1}=\Phi_{G_1}^\varepsilon \circ \Phi_{G_2}^{\varepsilon^2}\,,
\label{eq:KGW_canonicalTransf}
\end{align} 
with $G_1$, $G_2$ two generating Hamiltonians and with times $\varepsilon$, $\varepsilon^2$ of different order in $\varepsilon$. This canonical transformation maps the original Hamiltonian $H_\varepsilon$ into a new Hamiltonian in the form:
\begin{align}
H_\varepsilon\circ \mathcal{C}^{-1}&=
e^{\varepsilon^2 L_{G_2}} e^{\varepsilon L_{G_1}} (H_0 + \varepsilon H_1)  \notag\\
&= (1 + \varepsilon^2 L_{G_2} + \mathcal{O}(\varepsilon^4))(1 + \varepsilon L_{G_1} + \frac{\varepsilon^2}{2}L_{G_1}^2 + \mathcal{O}(\varepsilon^3))(H_0 + \varepsilon H_1)  \notag\\
&= H_0 + \varepsilon \underbrace{(L_{G_1}H_0 + H_1)}_{Z_1} + \varepsilon^2 \underbrace{(L_{G_2}H_0 + L_{G_1}H_1 + \frac{1}{2}L_{G_1}^2 H_0)}_{Z_2} + \mathcal{O}(\varepsilon^3)\,,
\label{eq:KGW_transformedHam}
\end{align}
where we used Lemma \ref{lem:flowComposition} to write the flow composition in algebraic form, expand the transformed Hamiltonian and collect the contributions based on the perturbative parameter $\varepsilon$. We can thus denote by $Z_1$, $Z_2$ the first- and second-order terms in the transformed Hamiltonian. Recalling the antisymmetry of the Poisson brackets, implying $L_{G_n}H_0 = -L_{H_0}G_n$, we set the so called first- and second-order homological equations:
\begin{align}
    -L_{H_0}G_1 + H_1 &= Z_1\,, 
    \label{eq:KGW_firstOrdHomol}\\
    -L_{H_0}G_2 + F_2 &= Z_2 \,,
    \qquad
    F_2 := L_{G_1}H_1 + \frac{1}{2}L_{G_1}^2H_0\,,
    \label{eq:KGW_secondOrdHomol}
\end{align}
having the generating Hamiltonians $G_1$, $G_2$ and the first- and second-order corrections $Z_1$, $Z_2$ as the unknowns. 

Observe that the generating Hamiltonians $G_1$, $G_2$ enter the equations only through the action of the $L_{H_0}$ operator. This means that they are defined up to terms in the kernel of $L_{H_0}=\{\cdot,H_0\}$. Thus, for example, we are free to add terms depending only on $|\psi|^2$, due to the gauge invariance of the unperturbed equations. This freedom can be used to require that $G_1$, $G_2$ satisfy specific properties, such as having zero average with respect to the flow of $h$, i.e. $\flowavg{G_1}=0$ and $\flowavg{G_2}=0$.

Computing the normal form thus reduces to solving the two homological equations \eqref{eq:KGW_firstOrdHomol} and \eqref{eq:KGW_secondOrdHomol}. Observe that, due to the distinct time orders $\varepsilon$, $\varepsilon^2$ in the canonical transformation \eqref{eq:KGW_canonicalTransf}, the two homological equations are hierarchically decoupled: the first-order homological equation \eqref{eq:KGW_firstOrdHomol} sets $G_1$, $Z_1$ and, once they are known, the second-order homological equation \eqref{eq:KGW_secondOrdHomol} allows to compute $G_2$, $Z_2$.
We can therefore focus on the two homological equations separately, adapting the standard averaging principle to our case to derive the desired unknowns.

\subsection{First order correction $Z_1$}
In order to find the first unknown $Z_1$, we compute the average of the first-order homological equation \eqref{eq:KGW_firstOrdHomol} along the flow of $h$. This allows to impose the definition \ref{def:normalForm} of normal form: we require $\{Z_1, h\} = 0$, which implies that $Z_1$ is invariant with respect to the average flow of $h$, i.e. $\flowavg{Z_1} = Z_1$. We thus have:
\begin{equation}
-\flowavg{L_{H_0}G_1} + \flowavg{H_1 } =  Z_1 \,.
\end{equation}
Then, we elaborate on the first term of the left-hand side to show that it vanishes. We write explicitly the definition of time average, recalling that the unperturbed flow of $h$ is periodic of period $T=2\pi$. We use the identity $L_{H_0} = L_h + L_k$ and we recall that the operators $L_h$, $L_k$ commute. The Poisson bracket $L_k$ is independent from the time $s$ of the flow of $h$, so it can be brought outside the integral: 
\begin{align}
\flowavg{L_{H_0}G_1} =
\frac{1}{2\pi}\int_0^{2\pi} e^{sL_h}L_{H_0}G_1 \,\dint{s}
= \frac{1}{2\pi}\int_0^{2\pi} e^{sL_h}L_hG_1 \,\dint{s} +\frac{L_k}{2\pi}\int_0^{2\pi} e^{sL_h}G_1 \,\dint{s}\,.
\end{align}
In the last integral we recognize the average of $G_1$ along the flow of $h$. We exploit the freedom we have in defining $G_1$ to require it has vanishing average: $\flowavg{G_1}=0$. The last integral is therefore zero. 
As for the first integral, observe that we can reconstruct a total derivative in the integrand and trivially solve the integral, which vanishes by periodicity:
\begin{align}
\frac{1}{2\pi}\int_0^{2\pi} e^{sL_h}L_hG_1 \,\dint{s} =
\frac{1}{2\pi} \int_0^{2\pi} 
 \totderiv{}{s}\left( e^{sL_h}G_1\right)  \,\dint{s}
 = \frac{1}{2\pi} \left[ e^{sL_h}G_1\right]_0^{2\pi}=0\,.
\end{align}
The homological equation, averaged on the flow of $h$, thus reduces to:
\begin{equation}
Z_1 = \flowavg{H_1}\,,
\label{eq:Z1def_fistOrder}
\end{equation}
a result which is analogous to the standard averaging principle, but involves only the average on the flow of $h$ rather than on that of the complete $H_0$. The first-order normal form correction $Z_1$ is thus equal to the original first-order term $H_1$ averaged along the flow $\Phi_h^s$ of the $h$ contribution in the unperturbed Hamiltonian. 

\subsubsection{Explicit computations in the KGW case}
To perform the explicit computation of $Z_1 = \flowavg{H_1}$ we use coordinates $(\psi,\psibar,\phi,p_\phi)$. Recall that this choice is arbitrary, since the normal form procedure is independent from the specific coordinates used for computations. 
In our case, the result is straightforward: it is sufficient to recall that the flow $\Phi_h^s$ acts only on $\psi, \psibar$ by shifting the phase and that the only non vanishing terms in the integration are those invariant under phase shifting, which remain unchanged. 
As a result, recalling the definition of $H_1$ and denoting $(\nabla \psi)^2:= \nabla \psi \cdot \nabla\psi$, we have:
\begin{align*}
    H_1 :&= \int \left( \frac{|\nabla\phi|^2}{2} + \frac{|\nabla\psi+\nabla\psibar|^2}{4} + \frac{\phi}{2}(\psi+\psibar)^2 \right)\\
    &= \int \left( \frac{|\nabla\phi|^2}{2} + \frac{|\nabla\psi|^2}{2} + \phi|\psi|^2 \right) + \int \left( \frac{(\nabla \psi)^2 + (\nabla \psibar)^2}{4} + \frac{\phi}{2}(\psi^2+\psibar^2)\right)\,,
\end{align*}
from which it is immediate to derive:
\begin{align*}
    Z_1 = \flowavg{H_1} = \int \frac{|\nabla\phi|^2}{2} + \frac{|\nabla\psi|^2}{2} + \phi|\psi|^2 \,.
\end{align*}
This gives the first-order correction in the Hamiltonian normal form \ref{def:normalForm} and defines one of the two unknowns of the first-order homological equation \eqref{eq:KGW_firstOrdHomol}.

\subsection{First order generating Hamiltonian $G_1$}
To find the second unknown of the first-order homological equation \eqref{eq:KGW_firstOrdHomol}, 
we substitute in it the result \eqref{eq:Z1def_fistOrder} obtained for $Z_1$ and we reconstruct the deviation of $H_1$ from its average along the flow of $h$:
\begin{align*}
L_{H_0}G_1 &= H_1 - Z_1 = H_1 - \flowavg{H_1} =: \diff{H_1}\,.
\end{align*}
In order to solve for $G_1$, we need to invert the $L_{H_0}$ operator. Let us first proceed formally, in analogy to what observed in the standard averaging principle procedure. In order to exploit the structure of our problem, we use $L_{H_0} = L_h + L_k$ and collect the $L_k$ contribution, before formally inverting to solve for $G_1$:
\begin{gather}
    L_h(1 + L_h^{-1}L_k)G_1 = \diff{H_1}\,, \notag\\
    G_1 = (1 + L_h^{-1}L_k)^{-1}L_h^{-1}\diff{H_1}\,.
\end{gather}
Observe that, from Lemma \ref{lem:eqLhG}, we already have a definition for the $L_h^{-1}$ operator. 
In order to compute $G_1$, we thus need to understand how to describe the action of the operator $(1+L_h^{-1}L_k)^{-1}$. To this aim, let us proceed by formally expanding it in series:
\begin{align}
    (1+L_h^{-1} L_k)^{-1} = 1 - L_h^{-1} L_k + (L_h^{-1} L_k)^2 + \ldots
\label{eq:operatorSeries}
\end{align}
In this way, the problem reduces to compute successive (potentially infinite) applications of the operators $L_k$ and $L_h^{-1}$, which are already defined. 

\subsubsection{Explicit computations in the KGW case}
In order to perform explicit computations, let us focus again on coordinates $(\psi,\psibar,\phi,p_\phi)$. Using the definition of $H_1$ and the computed average $\flowavg{H_1}$ it is immediate to obtain:
\begin{align*}
    \diff{H_1} = H_1 - \flowavg{H}_1 = \int \frac{(\nabla\psi)^2 + (\nabla\psibar)^2}{4} + \frac{\phi(\psi^2+\psibar^2)}{2}\,.
\end{align*}
As a next step, we need to understand how to describe the action of the $L_h^{-1}$ and $L_k$ operators, to be iteratively applied to $\diff{H_1}$.
\paragraph{Action of $L_h^{-1}$.} 
    Consider first the $L_h^{-1}$ operator, which is a time average along the periodic flow $\Phi_h^s=(e^{-is}\psi_0,e^{is}\psibar_0)$ weighted by time $s$, as described in Lemma \ref{lem:eqLhG}. Observe that the composition with this flow modifies only the coordinates $\psi,\psibar$. Thus, if a given functional $F$ is polynomial in these coordinates, each term in $e^{sh}F = F \circ \Phi_h^s$ will depend on $s$ only through a phase $e^{i(-n_\psi+n_{\psibar})s} =:e^{\pm ins}$, with exponents $n_\psi$, $n_{\psibar}$ denoting the degree of $\psi$, $\psibar$ in that term, and $n:=|n_\psi -n_{\psibar}|$. The integration in time $s$ will therefore focus on the weighted phases $se^{\pm ins}$ only. For example, for $F=\diff{H_1}$:
    \begin{align*}
        L_h^{-1}\diff{H_1} &= \frac{1}{2\pi}\int_0^{2\pi} s e^{sL_h}\left[ \int  \frac{(\nabla\psi)^2}{4} + \frac{ (\nabla\psibar)^2}{4} + \frac{\phi \psi^2}{2} +  \frac{\phi\psibar^2}{2} \right]\dint{s}\\
        & = \frac{1}{2\pi} \int_0^{2\pi} \left[ \int se^{-i2 s} \frac{(\nabla\psi_0)^2}{4} + se^{+i2 s}\frac{ (\nabla\psibar_0)^2}{4} + se^{-i2 s}\frac{\phi \psi_0^2}{2} + se^{+i2 s} \frac{\phi\psibar_0^2}{2} \right]\dint{s}\\
        & = \int \left[ \left(\frac{1}{2\pi} \int_0^{2\pi} se^{-i2 s} \dint{s} \right)  \frac{(\nabla\psi_0)^2}{4} + \left(\frac{1}{2\pi} \int_0^{2\pi} se^{+i2 s} \dint{s} \right) \frac{ (\nabla\psibar_0)^2}{4} \right.\\
        &\,\,\,\,\,\,\,\,\,\,+ \left.
        \left(\frac{1}{2\pi} \int_0^{2\pi}
        se^{-i2 s}\dint{s} \right) \frac{\phi \psi_0^2}{2} \,\,+ 
        \left(\frac{1}{2\pi} \int_0^{2\pi} se^{+i2 s} \dint{s}\right) \frac{\phi\psibar_0^2}{2} \right]\,. 
    \end{align*}
    Integration in time $s$ thus involves only terms in the form:
    \begin{align*}
        \frac{1}{2\pi}\int_0^{2\pi} s e^{\pm ins}\dint{s}&= \frac{1}{2\pi}\left[ s \frac{e^{\pm ins}}{\pm in} \right]_0^{2\pi} - \frac{1}{2\pi}\int_0^{2\pi} \frac{e^{\pm ins}}{\pm in} ds
        = \mp \frac{i}{n} &&\text{for }n\neq0\,,\\
        \frac{1}{2\pi}\int_0^{2\pi} s e^{\pm ins}\dint{s}&=\frac{1}{2\pi}\int_0^{2\pi} s\dint{s} = \frac{1}{2\pi}\left[ \frac{s^2}{2}\right]^{2\pi}_0 =\pi 
        &&\text{for } n=0\,,
    \end{align*}
    which is easily solved through integration by parts in the case $n \neq0$.  
    Thus, the operator $L_h^{-1}$ acts on a functional $F$ polynomial in $(\psi,\psibar)$ by adding a proportionality factor to each term: $\pi$ for terms invariant under phase shifting ($n=0$), and $\pm i n^{-1}$ to other terms, where the exponent $n=|n_\psi+n_{\psibar}|$ accounts for the degrees $n_\psi,\,n_{\psibar}$ of $\psi,\,\psibar$ in that term. 
    Note that in our case the operator $L_h^{-1}$ is always applied to deviations from averages, which by definition have vanishing average with respect to the flow of $h$. Thus, the case $n=0$ never occurs.
    Schematically, in our case we only need:
\begin{itemize}
    \item for terms proportional to $\psi^{n_\psi}$, the integrand has a phase $e^{-in_\psi s}$ and integration yields a proportionality constant $ +i/n_\psi$;
    \item for terms proportional to $\psibar^{n_{\psibar}}$, the integrand has a phase $e^{in_{\psibar}s}$ and integration yields a proportionality constant $ -i/n_\psi$.
\end{itemize}
To conclude our example, the application $L_h^{-1}\diff{H_1}$ yields:
\begin{align*}
    L_h^{-1}\diff{H_1} &= 
    -\frac{1}{2i} \int \frac{(\nabla\psi)^2}{4} + \frac{1}{2i} \int \frac{(\nabla\psibar)^2}{4} - \frac{1}{2i} \int \frac{\phi \psi^2}{2} + \frac{1}{2i} \int \frac{\phi \psibar^2}{2}\\
    & = \int \frac{i}{8}((\nabla\psi)^2 - (\nabla\psibar)^2) + \frac{i}{4} \int \phi(\psi^2 - \psibar^2)\,, \\
\end{align*}
where the subscript $0$ for the initial conditions has been omitted  for simplicity. We'll adopt the same notation in the following.
\paragraph{Action of $L_k$.}
    Consider now the $L_k$ operator. Its action on a generic functional $F(\psi,\psibar,\phi,p_\phi)$ is:
    \begin{align*}
        L_k F:= \left\{F, \int \frac{p_\phi^2}{2}\right\} \,.
    \end{align*}
    Observe that, since $L_k$ involves only $p_\phi$, the only non-vanishing contributions in $F$ are those depending on $\phi$. In our case we just need:
    \begin{align*}
        \left\{\int\phi^n, \int \frac{p_\phi^2}{2}\right\} = \int n \phi^{n-1} p_\phi\,,
    \end{align*}
    which lowers the degree of $\phi$ and raises that of $p_\phi$. Observe that, if $F$ is a polynomial of degree $n$ in $\phi$, only $n$ applications of $L_k$ will give a non-vanishing result. Thus, in this case, only a finite number of contributions will actually matter in the series \eqref{eq:operatorSeries} representing the $(1+L_h^{-1}L_k)^{-1}$ operator. \\

With this information, we can actually compute the first-order generating Hamiltonian $G_1$ by successive applications of the operators $L_h^{-1}$, $L_k$:
\begin{align*}
    L_h^{-1} \diff{H_1} &= \int \frac{i}{8}((\nabla\psi)^2 - (\nabla\psibar)^2 ) + \frac{i}{4} \phi (\psi^2 - \psibar^2)\,,\\
    L_k \,\,L_h^{-1} \diff{H_1} &= \int \frac{i}{4} p_\phi (\psi^2 - \psibar^2)\,,\\
    L_h^{-1} L_k \,\,L_h^{-1} \diff{H_1} &= -\int \frac{1}{8} p_\phi (\psi^2 + \psibar^2)\,,\\
    L_k \,\,L_h^{-1} L_k\,\, L_h^{-1} \diff{H_1} &= 0\,.
\end{align*}

Therefore, we conclude the computation of the first-order generating Hamiltonian:
\begin{align*}
    G_1 &= (1+L_h^{-1} L_k)^{-1} L_h^{-1} \diff{H_1} \\
    &= (1-L_h^{-1} L_k) \,\,\,\,\,\,L_h^{-1} \diff{H_1} \\
    &= \int \left[\frac{i}{8}((\nabla\psi)^2 - (\nabla\psibar)^2 ) + \frac{i}{4} \phi (\psi^2 - \psibar^2) + \frac{1}{8}p_\phi (\psi^2 + \psibar^2) \right]\,.
\end{align*}

\subsection{Second order correction $Z_2$}
Similarly to the first-order, we average the second-order homological equation \eqref{eq:KGW_secondOrdHomol} along the flow $\Phi_h^s$ of the harmonic unperturbed Hamiltonian $h$. The computation is exactly as in the first-order, and the $\flowavg{L_{H_0}G_2}$ term vanishes by requiring $G_2$ to have zero mean along the flow of $h$. Thus, we have:
\begin{align*}
    Z_2 = \flowavg{F_2}\,.
\end{align*}
To simplify the computation of $\flowavg{F_2}$, let us first manipulate its definition, by plugging in the first-order homological equation:
\begin{gather*}
    L_{G_1}^2 H_0 = L_{G_1}(L_{G_1}H_0) = L_{G_1}(\flowavg{H_1} - H_1) \,,\\
    F_2 := L_{G_1} H_1 + \frac{1}{2}L_{G_1}^2 H_0 
    =\frac{1}{2}L_{G_1}H_1 + \frac{1}{2}L_{G_1}\flowavg{H_1}\,.
\end{gather*}
When averaging, the last term vanishes, since we required $G_1$ to have zero average along the flow of $h$:
\begin{align*}
    \flowavg{ L_{G_1}\flowavg{H_1} } =
    \flowavg{\{\flowavg{H_1}, G_1\}} = \{\flowavg{H_1}, \flowavg{G_1}\} = \{\flowavg{H_1}, 0\} = 0\,,
\end{align*}
and we can use the same trick in the remaining term, by adding and subtracting the average $\flowavg{H_1}$ to reconstruct $\diff{H_1}:= H_1-\flowavg{H_1}$:
\begin{align*}
\flowavg{F_2} &= \frac{1}{2} \flowavg{L_{G_1}H_1} \\
&= \frac{1}{2}\flowavg{\{H_1 - \flowavg{H_1}, G_1\}} + \frac{1}{2}\flowavg{\{\flowavg{H_1}, G_1\}}\\
&= \frac{1}{2}\flowavg{\{\diff{H_1}, G_1\}}\,.
\end{align*}
Calculating this is usually simpler than computing $F_2$. The latter, however, is still necessary if we want to compute the generating Hamiltonian $G_2$ as well. Thus, in this paper, we proceed by computing $F_2$ and then by averaging along $h$. 

\subsubsection{Explicit computations in the KGW case}
To obtain $F_2$ it is necessary to compute the Poisson brackets $L_{G_1}(H_1+\flowavg{H_1})=\{H_1+\flowavg{H_1},G_1\}$. The calculations are reported in Appendix \ref{sec:app_LG1PoissonBrackets} and the final result reads:
\begin{align*}
    F_2 =& \frac{1}{2} L_{G_1} (H_1 + \flowavg{H_1}) \\
    =& -\frac{1}{8} \int |\Delta \psi|^2 + 
    \frac{1}{4} \int \phi \left( \psibar \Delta \psi + \psi \Delta \psibar \right) 
    +\frac{i}{16}\int p_\phi \left[\psibar \Delta\psi-\psi\Delta\psibar\right] +\\
    &-\frac{1}{2} \int \phi^2 |\psi|^2 + \frac{1}{16} \int |\psi|^4 +\\
    &-\frac{1}{8} \int \left( (\Delta \psi)^2 + (\Delta \psibar)^2 \right) + \frac{1}{8} \int |\psi|^2 (\psi^2 + \psibar^2) + \frac{1}{2 \cdot 16} \int (\psi^4 + \psibar^4) +\\
    &-\frac{1}{8} \int (\Delta \phi) (\psi^2+ \psibar^2) + \frac{1}{2} \int \phi \left( \psi \Delta \psi + \psibar \Delta \psibar \right) - \frac{1}{2} \int \phi^2 (\psi^2 + \psibar^2) + \\
    &-\frac{i}{8} \int p_\phi (\psi \Delta \psi - \psibar \Delta \psibar) + \frac{i}{4} \int \phi p_\phi (\psi^2 - \psibar^2) \,.
\end{align*}
The second-order correction $Z_2$ is finally computed by averaging along the flow of $h$, which - we recall - acts only on $\psi, \psibar$ by shifting the phase, so that through integration it lets all terms vanish except for those invariant under phase shifting, which remain unchanged:
\begin{align*}
     Z_2 := \flowavg{F_2} =& -\frac{1}{8} \int |\Delta \psi|^2 + 
    \frac{1}{4} \int \phi \left( \psibar \Delta \psi + \psi \Delta \psibar \right) 
    +\frac{i}{16}\int p_\phi \left[\psibar \Delta\psi-\psi\Delta\psibar\right] +\\
    &- \frac{1}{2} \int \phi^2 |\psi|^2 + \frac{1}{16} \int |\psi|^4 \,.
\end{align*}

\subsection{Second order generating Hamiltonian $G_2$}
Let us finally compute the second-order generating Hamiltonian $G_2$. To this aim, we proceed as for the first-order, by inverting the homological equation and expanding in series the operator:
\begin{align*}
    G_2 = (1 - (L_h^{-1} L_k) + (L_h^{-1} L_k)^2 + \ldots)L_h^{-1}\diff{F_2}\,.
\end{align*}
We can proceed directly with the explicit computations.

\subsubsection{Explicit computations in the KGW case}
Let us first write the difference $\diff{F_2}:= F_2 - \flowavg{F_2}$:
\begin{align*}
    \diff{F_2} =& F_2 - \flowavg{F_2}\\
    =& -\frac{1}{8} \int \left( (\Delta \psi)^2 + (\Delta \psibar)^2 \right) + \frac{1}{8} \int |\psi|^2 (\psi^2 + \psibar^2) + \frac{1}{32} \int (\psi^4 + \psibar^4) +\\
    &-\frac{1}{8} \int (\Delta \phi) (\psi^2 + \psibar^2) + \frac{1}{2} \int \phi (\psi \Delta \psi + \psibar \Delta \psibar) - \frac{1}{2} \int \phi^2 (\psi^2 + \psibar^2) +\\
    &-\frac{i}{8} \int p_\phi (\psi \Delta \psi - \psibar \Delta \psibar) + \frac{i}{4} \int \phi p_\phi (\psi^2 - \psibar^2)\,.
\end{align*}
The remaining computation now consists in successive applications of the operators $L_h^{-1}$ and $L_k$, whose action has already been described in the first-order analysis. In this case, observe that $\diff{F_2}$ has at most a quadratic dependence on the $\phi$ coordinate, meaning that three successive applications of the $L_k$ operator will let all contributions vanish. The only relevant contributions are computed as follows:
\begin{gather*}
    \begin{aligned}
    L_h^{-1} \diff{F_2} =& -\frac{i}{16} \int \left( (\Delta \psi)^2 - (\Delta \psibar)^2 \right) + \frac{i}{16} \int |\psi|^2 (\psi^2 - \psibar^2) + \frac{i}{4 \cdot 32} \int (\psi^4 - \psibar^4) +\\
    &-\frac{i}{16} \int (\Delta \phi) (\psi^2 - \psibar^2) + \frac{i}{4} \int \phi \left( \psi \Delta \psi - \psibar \Delta \psibar \right) - \frac{i}{4} \int \phi^2 (\psi^2 - \psibar^2) +\\
    &+\frac{1}{16} \int p_\phi (\psi \Delta \psi + \psibar \Delta \psibar) - \frac{1}{8} \int \phi p_\phi (\psi^2 + \psibar^2)\,,
    \end{aligned}\\
    \begin{aligned}
    L_k L_h^{-1} \diff{F_2} =&
    \, \frac{i}{4} \int p_\phi (\psi \Delta \psi - \psibar \Delta \psibar) - \frac{i}{4} \int2 \phi p_\phi (\psi^2 - \psibar^2) - \frac{1}{8} \int p_\phi^2 (\psi^2 + \psibar^2) +\\
    &- \frac{i}{16} \int (\psi^2 - \psibar^2) \Delta p_\phi\,,
    \end{aligned}
    \\
    \begin{aligned}
    L_h^{-1} L_k L_h^{-1} \diff{F_2} =&
    -\frac{1}{8} \int p_\phi (\psi \Delta \psi + \psibar \Delta \psibar) + \frac{1}{4} \int \,\,\phi p_\phi (\psi^2 + \psibar^2) +\\
    &- \frac{i}{16} \int p_\phi^2 (\psi^2 - \psibar^2) 
    + \frac{1}{32} \int (\Delta p_\phi) (\psi^2 + \psibar^2)\,,
    \end{aligned}
    \\
    L_k L_h^{-1} L_k L_h^{-1} \diff{F_2} = \frac{1}{4} \int p_\phi^2 (\psi^2 + \psibar^2) \,,\\
    (L_h^{-1} L_k)^2 L_h^{-1} \diff{F_2} = \frac{i}{8} \int p_\phi^2 (\psi^2 - \psibar^2)\,.
\end{gather*}
The second-order generating Hamiltonian thus reads:
\begin{align*}
    G_2 =& (1+ L_h^{-1} L_k)^{-1} L_h^{-1} \diff{F_2} \\
    =& (1 - L_h^{-1} L_k + (L_h^{-1} L_k)^2) L_h^{-1} \diff{F_2}\\
    =& -\frac{i}{16} \int \left( (\Delta \psi)^2 - (\Delta \psibar)^2 \right) + \frac{i}{16} \int |\psi|^2 (\psi^2 - \psibar^2) + \frac{i}{4 \cdot 32} \int (\psi^4 - \psibar^4) +\\
    & -\frac{i}{16} \int (\Delta \phi) (\psi^2 - \psibar^2) + \frac{i}{4} \int \phi (\psi \Delta \psi - \psibar \Delta \psibar) - \frac{i}{4} \int \phi^2 (\psi^2 - \psibar^2) + \\
    & +\frac{3}{16} \int p_\phi (\psi \Delta \psi + \psibar \Delta \psibar) - \frac{3}{8} \int \phi p_\phi (\psi^2 + \psibar^2) \\
    & + \frac{3i}{16} \int p_\phi^2 (\psi^2 - \psibar^2) 
    - \frac{1}{32} \int (\Delta p_\phi) (\psi^2 + \psibar^2)\,. 
\end{align*}

\subsection{Remark on Higher Orders}
\label{sec:higherOrders}
Observe that the construction suggested in this paper can be extended to compute the normal form to higher orders. By composing the $(n-1)$-th order canonical transformation with a new flow $\Phi_{G_n}^{\varepsilon^n}$, the transformed Hamiltonian can be studied to construct the $n$-th order normal form. This is done by setting $n$-th homological equation:
\begin{align}
    -L_{H_0}G_n + F_n &= Z_n\,,
\label{eq:KGW_nthOrdHomol}
\end{align} 
with unknowns $G_n$, $Z_n$ and with the term $F_n$ depending only on known previous orders contributions. 
In analogy with the computations performed in this paper, the $n$-th order correction $Z_n$ can be obtained by averaging the $n$-th order homological equation \eqref{eq:KGW_nthOrdHomol}, resulting in:
\begin{align*}
    Z_n = \flowavg{F_n}\,,
\end{align*}
while the generating Hamiltonian $G_n$ is defined by substituting this result in the $n$-th order homological equation \eqref{eq:KGW_nthOrdHomol}, by reconstructing the difference $\diff{F_n}$ and by solving for $G_n$:
\begin{align*}
    G_n &= (1+ L_h^{-1}L_k)^{-1}L_h^{-1} \diff{F_n}\,,
    \qquad
    \diff{F_n} \equiv F_n - \flowavg{F_n}\,.
\end{align*}
As for the first- and second-order, the explicit computation of the inverse operator $(1+ L_h^{-1}L_k)^{-1}$ can be performed with the algebraic expansion:
\begin{align*}
    (1+L_h^{-1} L_k)^{-1} = 1 - L_h^{-1} L_k + (L_h^{-1} L_k)^2 + \ldots
\end{align*}
involving the iterated application of the $L_h^{-1}$, $L_k$ operators described in this paper. As already observed, when applying this operator to a functional $\diff{F_n}$ with a polynomial dependence on $\phi$ only a finite list of terms will result in a non-vanishing contribution. The series defining the inverse operator would thus reduce to a finite number of terms, allowing for an explicit computation of $G_n$.

\section{Conclusions}
\label{sec:discussionConclusions}
In this paper, we derive the Hamiltonian normal form for the KGW problem \eqref{eq:KG-W}, to second order, in a regime where the parameter $\mu^2$ is large. 
We write the Hamilton equations obtained by neglecting the remainders and analyze the first- and second-order approximations of the KGW system in a normal form sense. We observe that a SW system emerges as a first-order approximation, which can be further approximated by a SP equation in the limit $\varepsilon \to 0$, provided that the time oscillations of the complex massive field ($\Psi$) are not too large. We also provide the explicit second-order corrections to the SW system.
Our constructive procedure - a modification of the standard averaging principle that admits a component with unbounded flow in the unperturbed Hamiltonian - could be iteratively applied to compute higher-order approximations.

The perturbative scheme used here preserves, by construction, the $L^2$ norm of the complex massive field 
to all orders. Such a result is obviously relevant from a physical point of view. For example, in the dark matter problem, it corresponds to the conservation of the total dark mass in the system to any perturbative order. In turn, this means that, with respect to the original KGW dynamics, the total mass is approximately conserved over long time scales ($1/\varepsilon^n$ to order $n$), even when the original dynamics and the approximating one are far apart from each other (which typically happens on the time scale $1/\varepsilon$). Another important point concerns the possibility to explain the cooling mechanism of dark matter, i.e. why, starting with a relativistic dynamics, particles of extremely small mass that interact gravitationally move according to approximate, non relativistic equations (as the SW and the SP equations). 

We observe that interpreting the KGW system as a set of coupled harmonic oscillators and free particles proves useful when studying the analytical details of the normal form procedure. Moreover, such an idea could be applied to physical systems having nothing to do with those considered here, but that can be reduced - from a mathematical point of view - to normal modes coupled with free particles.


\section*{Acknowledgements}

We acknowledge GNFM (INdAM), the Italian national group for Mathematical Physics.

\noindent We acknowledge ICSC – Centro Nazionale di Ricerca in High Performance Computing, Big Data and Quantum Computing, funded by European Union – NextGenerationEU.

\bibliography{bibliography}

\clearpage
\appendix

\section{A KGW system modeling ultralight scalar dark matter}
\label{sec:app_phys}

A quantum field theory model of dark matter would involve a massive, real scalar field $U$ and a mediator, massless scalar field $\Phi$. The role of the latter field is crucial: all known interactions in fundamental physics are mediated by certain particle fields. Since dark matter particles are supposed to interact only gravitationally, $\Phi$ plays the role of a pseudo-graviton field. Assuming $\Phi$ to be a scalar and not a tensor field is a drastic simplification. A further simplification consists in treating the fields as classical ones, i.e. as functions. Quantum effects enter such an effective theory only through the fundamental physical constants. The equations of the effective theory consist in a Klein-Gordon equation for $U$, and a massless Klein-Gordon or wave equation for $\Phi$, plus coupling terms. After some inspection, one realizes that the minimal choice of Yukawa-like coupling gives the correct physics in the stationary case.
The system in physical units is the following:
\begin{equation}
\label{eq:KGWphys}
\left\{ \begin{array}{l} 
\Box U =\frac{m^2c^2}{\hbar^2}U+2\frac{m^2}{\hbar^2}\Phi U \\ 
\Box \Phi=4\pi Gm U^2
\end{array}\right.\ ;\ \Box=\Delta_X-c^{-2}\partial_T^2\ .
\end{equation} 
Looking for the stationary solutions of this system, the second equation becomes the Poisson equation $\Delta\Phi=4\pi GmU^2$, yielding the gravitational field due to a matter distribution with density $U^2$. The $N$-particle theory then requires a normalization condition of the form $\int U^2\dint{^3}X=N$. The choice of the coupling term in the first equation is then dictated by the requirement that the coupling has a Lagrangian or Hamiltonian form. 

Equations \eqref{eq:KGWphys} can be suitably rescaled in such a way to get a theory depending on a single control parameter. Indeed, by the rescaling
\begin{eqnarray}
&& X=\lambda x\ ;\ T=\frac{\lambda}{c} t\ ;\\
&& U=\sqrt{\frac{N}{4\pi}}\lambda^{-3/2}u\ ;\ \Phi=\frac{GNm}{\lambda}\phi\ , 
\end{eqnarray}
and choosing 
\begin{equation}
\label{eq:lambda}
\lambda=\frac{\hbar^2}{GNm^3}\ ,
\end{equation}
system \eqref{eq:KGWphys} cleans up into the dimensionless form
\begin{equation}
\label{eq:KGWdimless}
\left\{ \begin{array}{l} 
\Box u =\mu^2u+2\phi u \\ 
\Box \phi=u^2 
\end{array}\right.\ ,
\end{equation} 
with the parameter $\mu^2$ expressed as
\begin{equation}
\label{eq:mu2}
\mu^2\equiv\frac{1}{N^2}\left(\frac{m_P}{m}\right)^4\ ,
\end{equation}
$m_P=\sqrt{c\hbar/G}$ being the Planck mass. The rescaled normalization condition of the initial datum reads 
$\|u(0)\|^2=\int u^2(0,x)\dint{^3}x=4\pi$. Now, since $\|u(t)\|^2$ is not preserved by the dynamics of the system, the exact value $4\pi$ obtained above has no particular meaning. Any number can be chosen in place of it, for example one (however, with a further accurate rescaling one may reduce the $4\pi$ value to one exactly). 

The system \eqref{eq:KGWdimless} coincides with the KGW system \eqref{eq:KG-W} studied in this paper, and the expression of the parameter \eqref{eq:mu2} coincides with that given in the text, formula \eqref{eq:mu}. 

\medskip

\section{Computation of $F_2$}
\label{sec:app_LG1PoissonBrackets}
We report here the calculations needed to compute:
\begin{align*}
    F_2 =& \frac{1}{2} L_{G_1} (H_1 + \flowavg{H_1}) \,.
\end{align*}
To begin with, let us report here the definitions we have from the first-order:
\begin{align*}
H_1 &= \int \frac{(\nabla\phi)^2}{2} + \frac{|\nabla\psi|^2}{2} + \phi|\psi|^2 + \int \frac{(\nabla\psi)^2 + (\nabla\psibar)^2}{4} + \frac{\phi}{2}(\psi^2 + \psibar^2)\\
\flowavg{H_1} &= \int \frac{(\nabla\phi)^2}{2} + \frac{|\nabla\psi|^2}{2} + \phi|\psi|^2\\
H_1+\flowavg{H_1} &= \int (\nabla\phi)^2 + |\nabla\psi|^2 + 2\phi|\psi|^2 + \int \frac{(\nabla\psi)^2+(\nabla\psibar)^2}{4} + \frac{\phi}{2}(\psi^2+\psibar^2)\\
G_1 &= \int \frac{i}{8}\left((\nabla\psi)^2-(\nabla\psibar)^2\right) + \frac{i}{4}\phi(\psi^2-\psibar^2) + \int\frac{p_\phi}{8}(\psi^2+\psibar^2)
\end{align*}
We are now ready to compute the required Poisson brackets. 
We obtain: 
\begingroup
\allowdisplaybreaks
\begin{align*}
    L_{G_1}&(H_1+\flowavg{H_1}) =\\
    =& \,\,\,\,\,\,\,\{H_1+\flowavg{H_1}, G_1\}\\
    =& \,\,\,\,\,\,\,\left\{\int(\nabla\phi)^2, \frac{1}{8}\int p_\phi (\psi^2+\psibar^2)\right\} +\\
    &+ \left\{\int|\nabla\psi|^2, \frac{i}{8}\int\left((\nabla\psi)^2-(\nabla\psibar)^2\right) + \frac{i}{4}\int\phi(\psi^2-\psibar^2) + \int\frac{p_\phi}{8}(\psi^2+\psibar^2)\right\} +\\
    &+ \left\{2\int\phi|\psi|^2, \frac{i}{8}\int\left((\nabla\psi)^2-(\nabla\psibar)^2\right) + \frac{i}{4}\int\phi(\psi^2-\psibar^2) + \int\frac{p_\phi}{8}(\psi^2+\psibar^2)\right\} +\\
    &+ \left\{\frac{1}{4}\int(\nabla\psi)^2, -\frac{i}{8}\int(\nabla\psibar)^2 - \frac{i}{4}\int\phi\psibar^2 + \int\frac{p_\phi}{8}\psibar^2\right\} +\\
    &+ \left\{\frac{1}{4}\int(\nabla\psibar)^2, \frac{i}{8}\int(\nabla\psi)^2 + \frac{i}{4}\int\phi\psi^2 + \int\frac{p_\phi}{8}\psi^2\right\} +\\
    &+ \left\{\frac{1}{2}\int\phi\psi^2, \frac{i}{8}\int\left((\nabla\psi)^2-(\nabla\psibar)^2\right) + \frac{i}{4}\int\phi(\psi^2-\psibar^2) + \int\frac{p_\phi}{8}(\psi^2+\psibar^2)\right\} +\\
    &+ \left\{\frac{1}{2}\int\phi\psibar^2, \frac{i}{8}\int\left((\nabla\psi)^2-(\nabla\psibar)^2\right) + \frac{i}{4}\int\phi(\psi^2-\psibar^2) + \int\frac{p_\phi}{8}(\psi^2+\psibar^2)\right\} \\
\end{align*}
\begin{align*}
    L_{G_1} &(H_1 + \flowavg{H_1}) =\\
    =& \,\,\,\,\,\,\,\, \frac{1}{8}\int(-2\Delta\phi)(\psi^2+\psibar^2) +\\
    &+ (-i)\left[\frac{i}{8}\int\left[-(\Delta\psibar)(2\Delta\psibar) - (\Delta\psi)(2\Delta\psi)\right] + \right.\\
    &\left. + \frac{i}{4}\int\phi\left[(\Delta\psibar)(2\psibar) + (\Delta\psi)(2\psi)\right] +\frac{1}{8}\int p_\phi\left[-(\Delta\psibar)(2\psibar) + (\Delta\psi)(2\psi)\right]\right] +\\
    &+ (-i)\left[\frac{2i}{8}\int\phi\left[\psibar(2\Delta\psibar) + \psi(2\Delta\psi)\right] + \frac{2i}{4}\int\phi^2\left[-\psibar(2\psibar) - \psi(2\psi)\right] +\right.\\
    &\left. + \frac{2}{8}\int \phi p_\phi\left[\psibar(2\psibar) - \psi(2\psi)\right]\right] + \frac{2}{8}\int|\psi|^2(\psi^2+\psibar^2) +\\
    &+ (-i)\left[-\frac{i\cdot2}{4\cdot8}\int(2\Delta\psi)(2\Delta\psibar) 
    - \frac{i}{4\cdot4}\int \phi\left[-(2\Delta\psi)(2\psibar)-(2\Delta\psibar)(2\psi)\right] 
    \right. \\
    &\left.
    + \frac{1}{4\cdot8}\int p_\phi\left[-(2\Delta\psi)(2\psibar)+(2\Delta\psibar)(2\psi) \right]
    \right] +\\      
    &+ (-i)\left[
    \frac{i}{2\cdot8}\int\phi \left[(2\psi)(2\Delta\psibar)+(2\psibar)(2\Delta\psi) \right]
    + \frac{i\cdot2}{2\cdot4}\int[-\phi^2(2\psi)(2\psibar)] + \right.\\
    &+\left. \frac{1\cdot0}{2\cdot8}\int\phi p_\phi(2\psi)(2\psibar)
    \right] + \left[\frac{1}{2\cdot8}\int\psi^2(\psi^2+\psibar^2) + \frac{1}{2\cdot8}\int\psibar^2(\psi^2+\psibar^2)\right] \\
\end{align*}
\begin{align*}
    L_{G_1} &(H_1 + \flowavg{H_1}) =\\
    =& -\frac{1}{4} \int (\Delta \phi) (\psi^2 + \psibar^2) + \\
    &-\frac{1}{4} \int \left( (\Delta \psi)^2 + (\Delta \psibar)^2 \right) + \frac{1}{2} \int \phi \left( \psi \Delta \psi + \psibar \Delta \psibar \right) - \frac{i}{4} \int p_\phi \left( \psi \Delta \psi - \psibar \Delta \psibar \right) + \\
    &+ \frac{1}{2} \int \phi (\psi \Delta \psi + \psibar \Delta \psibar) - \int \phi^2 (\psi^2+ \psibar^2) + \frac{i}{2} \int \phi p_\phi (\psi^2 - \psibar^2) +\\
    &+ \frac{1}{4} \int |\psi|^2 (\psi^2+ \psibar^2) 
    -\frac{1}{4} \int |\Delta \psi|^2 
    + \frac{1}{2} \int \phi \left[ \psibar \Delta \psi + \psi \Delta \psibar \right] +\\
    &+\frac{i}{8}\int p_\phi \left[\psibar \Delta\psi-\psi\Delta\psibar\right]
    - \int \phi^2 |\psi|^2 + \frac{1}{16} \int (\psi^2 + \psibar^2)^2 \,.
\end{align*}
\endgroup

Thus, collecting first the terms invariant under phase shifting, $F_2$ reads:
\begin{align*}
    F_2 =& \frac{1}{2} L_{G_1} (H_1 + \flowavg{H_1}) \\
    =& -\frac{1}{8} \int |\Delta \psi|^2 + 
    \frac{1}{4} \int \phi \left( \psibar \Delta \psi + \psi \Delta \psibar \right) 
    +\frac{i}{16}\int p_\phi \left[\psibar \Delta\psi-\psi\Delta\psibar\right]+\\
    &-\frac{1}{2} \int \phi^2 |\psi|^2 + \frac{1}{16} \int |\psi|^4 +\\
    &-\frac{1}{8} \int \left( (\Delta \psi)^2 + (\Delta \psibar)^2 \right) + \frac{1}{8} \int |\psi|^2 (\psi^2 + \psibar^2) + \frac{1}{2 \cdot 16} \int (\psi^4 + \psibar^4) +\\
    &-\frac{1}{8} \int (\Delta \phi) (\psi^2+ \psibar^2) + \frac{1}{2} \int \phi \left( \psi \Delta \psi + \psibar \Delta \psibar \right) - \frac{1}{2} \int \phi^2 (\psi^2 + \psibar^2) + \\
    &-\frac{i}{8} \int p_\phi (\psi \Delta \psi - \psibar \Delta \psibar) + \frac{i}{4} \int \phi p_\phi (\psi^2 - \psibar^2) \,.
\end{align*}

\end{document}